\documentclass[prl,aps,showpacs,twocolumn,floatfix]{revtex4}
\usepackage{epsfig} \usepackage{graphics} \usepackage{bm}
\usepackage{amssymb}
\usepackage{graphicx}
\begin{document}

\preprint{Lebed-PRB}

\title{Orbital Effect for the Fulde-Ferrell-Larkin-Ovchinnikov Phase
in a Quasi-Two-Dimensional Superconductor in a Parallel Magnetic
Field}

\author{A.G. Lebed$^*$}

\affiliation{Department of Physics, University of Arizona, 1118 E.
4-th Street, Tucson, AZ 85721, USA}

\begin{abstract}
We theoretically study the orbital destructive effect against
superconductivity in a parallel magnetic field in the
Fulde-Ferrell-Larkin-Ovchinnikov (FFLO or LOFF) phase at zero
temperature in a quasi-two-dimensional (Q2D) conductor. We
demonstrate that at zero temperature a special parameter, $\lambda
= l_{\perp}(H)/d$, is responsible for strength of the orbital
effect, where $l_{\perp}(H)$ is a typical "size" of the
quasi-classical electron orbit in a magnetic field and $d$ is the
inter-plane distance. We discuss applications of our results to
the existing experiments on the FFLO phase in the organic Q2D
conductors $\kappa$-(ET)$_2$Cu(NCS)$_2$ and
$\kappa$-(ET)$_2$Cu[N(CN)$_2$]Cl.

\end{abstract}


\maketitle

It is well known that the orbital effect of electron motion in
external magnetic field destroys superconductivity [1]. In singlet
type-II superconductors, superconductivity is usually destroyed by
magnetic fields higher than the so-called upper critical field,
$H_{c2}$. For a 3D isotropic case, at zero temperature $H_{c2}(0)$
was calculated in Ref.[2], whereas temperature dependence of the
upper critical field, $H_{c2}(T)$, was found several years later
[3]. As to triplet superconductivity, it can be restored in some
cases in magnetic fields much higher than the $H_{c2}(0)$, as
theoretically predicted for quasi-one-dimensional (Q1D) [4,5],
quasi-two-dimensional (Q2D) [6], and isotropic 3D [7]
superconductors.

Note that superconductivity in singlet superconductors can also be
destroyed by spin effects, as was first demonstrated in Refs.
[8,9] (i.e., above the so-called Clogston-Chandrasekhar
paramagnetic limit, $H_P$). Nevertheless, Larkin, Ovchinnikov,
Felde, and Ferrell (LOFF) stressed [10,11] that situation with the
above mentioned paramagnetic destruction of superconductivity is
not so simple. Indeed, they showed that there might exist the FFLO
(or LOFF) superconducting non-uniform phase in restricted area of
magnetic fields, $H_p < H < H_{FFLO}$. This happens when the
orbital effect is small enough, which is realized in Q1D
superconductors in an arbitrary oriented magnetic field and in Q2D
superconductors for a magnetic field parallel the conducting
layers. As was shown in Ref.[12], the FFLO phase was stable in a
pure 1D case for arbitrary strong magnetic field in the absence of
the orbital effect. In Refs. [4,5,13,14], a possibility of the
FFLO phase to exist in real Q1D materials from chemical family
(TMTSF)$_2$X (X = ClO$_4$, PF$_6$, etc.) was studied taking into
account the orbital effect in a perpendicular magnetic field. In
Ref.[15], it was shown that the FFLO phase has to exist in the Q1D
superconductor (TMTSF)$_2$ClO$_4$, despite the orbital effect in a
parallel magnetic field. Some important signatures of the possible
existence of the FFLO phase were experimentally observed in
perpendicular [16,17] and parallel [18,19] magnetic fields in the
Q1D superconductors (TMTSF)$_2$ClO$_4$ and (TMTSF)$_2$PF$_6$.

As mentioned before, the second convenient case for a possible
observation of the FFLO phase is a Q2D superconductor in a
parallel magnetic field. In the absence of the orbital effect
(i.e., for a pure 2D case), such problem was considered in
Refs.[20-26] and some others (see for the references, reviews
[23,24]). The orbital effect was first considered in Refs. [27,28]
for high enough temperatures, $T \gg t_{\perp}$, for a Q2D
superconductor with $t_{\perp} \ll T_{c0}$. Here, $t_{\perp}$ is
the overlapping integral of electron wave functions, corresponding
to electron jumping in a perpendicular to the conducting planes
direction, $T_{c0}$ is superconducting transition temperature in
the absence of a magnetic field.  The main result of Refs.[27,28]
is that the orbital effect is of the relative order of
$t^2_{\perp}/T^2_{c0} \ll 1$. From experimental side, a plenty of
experimental works on Q2D organic and some other superconductors
have been performed [29-41] to establish the possible existence of
the FFLO phase in a parallel magnetic field.

The goal of our paper is to consider the orbital effect in a
parallel magnetic field in a Q2D conductor at zero temperature, in
contrast to Refs.[27,28]. We show that there exists new parameter,
$\lambda = l_{\perp}(H)/d$, where $l_{\perp}(H)$ is a typical
"size" of the quasi-classical electron trajectory in a magnetic
field and $d$ is the inter-layer distance. We show that $\lambda$
defines how many conducting layers participate in the creation of
one superconducting pair. In particular, we demonstrate that if
this parameter is small, then we have effectively the
superconducting pairing within almost one conducting layer and can
disregard the orbital effect. On the contrary, if this parameter
is large, then the superconducting pair is larger than the
inter-layer distance and it is necessary to take into account the
orbital effect against superconductivity. We compare the obtained
results with the existing experiments on the FFLO phase at low
temperatures in $\kappa$-(ET)$_2$Cu(NCS)$_2$, where the FFLO phase
has been the most firmly established [40].

Below, we consider a layered superconductor with the following Q2D
electron spectrum:
\begin{equation}
\epsilon({\bf p})= \epsilon_{\parallel} (p_x, p_y) + 2 t_{\perp}
\cos(p_z d) \ ,
\ \ \ t_{\perp} \ll \epsilon_F \ ,
\end{equation}
in a parallel magnetic field,
\begin{equation}
{\bf H} = (0,H,0) \  ,  \ \ \ \ {\bf A} = (0,0,-Hx) \ ,
\end{equation}
where $\hbar \equiv 1$.
Here, in-plane electron energy
$\epsilon_{\parallel} (p_x,p_y) \sim \epsilon_F$ with $\epsilon_F$
being the Fermi energy. Note that near 2D Fermi surface (FS),
\begin{equation}
\epsilon_{\parallel} (p_x,p_y)=\epsilon_F ,
\end{equation}
the Q2D electron spectrum (1) can be linearized:
\begin{equation}
\epsilon({\bf p}) - \epsilon_F =  v_x(p_y) [p_x - p_x(p_y)]
+ 2 t_{\perp} \cos(p_z d) \ ,
\end{equation}
where $v_x(p_y) = \partial \epsilon_{\parallel} (p_x,p_y) /
\partial p_x$ is a velocity component and $p_x(p_y)$ is the Fermi
momentum component along $x$ axis.

First, let us consider a qualitative physical picture of
superconducting pairing in the magnetic field (2) and study the
quasi-classical electron motion in the field. For simplicity, we
employ an isotropic in-plane electron spectrum with
\begin{equation}
\epsilon(p_x,p_y) = \frac{(p^2_x+p^2_y)}{2m}.
\end{equation}
For electrons with spectrum (4),(5) the second Newton's law can be
written in the magnetic field (2) as
\begin{equation}
\frac{dp_z}{dt}=\biggl(\frac{e}{c} \biggl)v_F H \sin \alpha ,
\end{equation}
where $\alpha$ is angle between the magnetic field and electron
position on the 2D FS (5). Note that electron velocity in
perpendicular to the conducting planes direction can be written as
$v_z(p_y) = 2t_{\perp} d \sin(p_zd)$ from Eq.(1). Therefore,
electron oscillates in time in the perpendicular direction in the
following way:
\begin{equation}
z = z_0 + \frac{2t_{\perp} d}{\omega_c \sin \alpha} \sin(\omega_c
\sin \alpha t), \ \ \omega_c= \frac{ev_F dH}{c} ,
\end{equation}
where
\begin{equation}
l_{\perp}(H) = \frac{\lambda d}{\sin \alpha}, \ \ \lambda =
\frac{4t_{\perp} }{\omega_c}
\end{equation}
is a typical "size" of electron orbit in the magnetic field (2).
From Eq.(8) it is directly seen that, at
\begin{equation}
\lambda \ll 1 ,
\end{equation}
the most electrons are localized on conducting planes. This means
 that the orbital effect is small and, under this
condition, we can expect that electrons form almost 2D
superconducting pairs. Therefore, the FFLO phase is expected to
survive.

Let us now consider quantitative quantum problem of the FFLO phase
formation in the presence of the orbital effect against
superconductivity.
 To obtain electron Hamiltonian in the magnetic
field (2), $\hat H(x;p_y,p_z,s)$, we make use of the Peierls
substitution method in Eq.(4) in the following way [6]:
\begin{equation}
p_x \rightarrow - i \biggl( \frac{d}{dx} \biggl),  \ \ p_z
\rightarrow p_z + \biggl( \frac{e}{c} \biggl) H x,
\end{equation}
where $s = \pm \frac{1}{2}$ is electron spin projection along
quantization $y$ axis.

Under such conditions, the Green's functions of the Q2D electrons
(4) in the magnetic field (2) obey the following differential
equation [42],
\begin{eqnarray}
[i \omega_n - \hat H(x;p_y,p_z,s)]G(i \omega_n; x, x_1;p_y,p_z;s)
= \delta(x-x_1) ,
\nonumber\\
 \biggl\{ i \omega_n - v_x (p_y) \biggl[ - i \frac{d}{dx} -
p_x(p_y) \biggl] +2 t_{\perp} \cos \biggl( p_z d + \frac{eHdx}{c}
\biggl)
\nonumber\\
+ 2 \mu_B H s \biggl\} G(i \omega_n; x, x_1;p_y,p_z;s) = \delta
(x-x_1) \ .
\end{eqnarray}
In Eq.(11), $\omega_n$ is the so-called Matsubara frequency [42]
and $\mu_B$ is the Bohr magneton. Let us solve Eq.(11)
analytically. As a result, for the Green's functions we obtain
\begin{eqnarray}
G (i \omega_n; x, x_1;p_y,p_z;s) = - i \frac{ sgn \
\omega_n}{v_x(p_y)}
 \exp \biggl[ - \frac{\omega_n (x-x_1)}{v_x(p_y)} \biggl]
\nonumber\\
\times \exp \biggl\{ \frac{ i \lambda(p_y)}{2} \biggl[ \sin
\biggl( p_zd + \frac{eHdx}{c} \biggl) - \sin \biggl( p_zd +
\frac{eHdx_1}{c} \biggl) \biggl] \biggl\}
\nonumber\\
\times \exp[i p_x(p_y)(x-x_1)] \exp \biggl[ \frac{2 i \mu_B s H
(x-x_1)}{v_x(p_y)} \biggl] ,
\end{eqnarray}
where $\lambda(p_y) = 4t_{\perp}c/eHdv_x(p_y)$.

To determine superconducting transition temperature as a function
of a magnetic field, $T_c(H)$, we derive the so-called Gor'kov's
equations [42] for the case of non-uniform superconductivity [4].
As a result, we obtain
\begin{eqnarray}
\Delta(x) = U \oint  \frac{d l}{v_{\perp}(l)} \int^{\infty}_{
|x-x_1| > \frac{|v_x(l)|}{\Omega}} \frac{2 \pi T dx_1}{v_x(l)
\sinh \biggl[ \frac{ 2 \pi T |x-x_1|}{ v_x (l)} \biggl] }
\nonumber\\
\times J_0 \biggl\{ 2 \lambda (l)  \sin \biggl[ \frac{
e H d (x-x_1)}{2 c} \biggl] \sin \biggl[ \frac{ e H d (x+x_1)}{2
c} \biggl] \biggl\}
\nonumber\\
\times \cos \biggl[ \frac{2  \mu_B H (x-x_1)}{v_x (l)} \biggl] \
\ \Delta(x_1) \ ,
\end{eqnarray}
where integration in Eq.(13) is made along 2D contour,
$\epsilon_{\parallel} (p_x,p_y)=\epsilon_F$, $v_{\perp}(l)$ is a
velocity component perpendicular to the contour, $U$ is an
effective electron-electron interactions constant, $\Omega$ is a
cut-off energy, $J_0(...)$ is the zero-order Bessel function.
[Note that, for simplicity, Eq.(13) is derived for singlet
$s$-wave superconductors].

We point out that Eq.(13) is the most general one among the
existing equations to determine the parallel upper critical field
in a layered s-wave superconductor. As the limiting cases, it
contains Ginzburg-Landau and Lawrence-Doniach equations [42,43] as
well as quasi-classical equation similar to the gap Eq. of
Ref.[2]. In particular, it takes into account quantum effects of
electrons motion in a magnetic field - the Bragg reflections - and
related $3D \rightarrow 2D$ dimensional crossovers of electrons
[6], which move in the extended Brillouin zone in a parallel
magnetic field.

If we disregard the orbital effect in Eq.(13), it reduces to the
following form
\begin{eqnarray}
1 = U \oint  \frac{d l}{v_{\perp}(l)} \int^{\infty}_{ |z|
> \frac{|v_x(l)|}{\Omega}} \frac{2 \pi T dz}{v_x(l) \sinh
\biggl[ \frac{ 2 \pi T |z|}{ v_x (l)} \biggl] }
\nonumber\\
\cos \biggl[ \frac{2  \mu_B H z}{v_x (l)} \biggl] \  \ \cos[k_1z]
\ ,
\end{eqnarray}
defining the FFLO phase in a pure 2D case. Below, we consider the
situation, where in the absence of the orbital effect small
electron spectrum anisotropic effects fix the wave vector $k_1$ of
the FFLO phase [28] and, thus, we have the following solution
\begin{equation}
\Delta_0(x)=\cos(k_1x) .
\end{equation}
Here, we apply the in-plane magnetic field (2) perpendicular to
the wave vector $k_1$ of the FFLO phase. Our task is to determine
which fields can be considered as small ones and, thus, do not
destroy the FFLO phase. We consider in-plane electron spectrum
anisotropy to be large enough to fix the FFLO wave vector and to
be small enough to influence the orbital effect [28]. In other
words, in the presence of the orbital effects, we use the
following equation, obtained for the isotropic in-plane electron
spectrum (5):
\begin{eqnarray}
\Delta(x) = \frac{g}{2} \biggl< \int^{\infty}_{ |x-x_1|
> \frac{|sin \alpha|}{\Omega}} \frac{2 \pi T dx_1}{v_F \sin \alpha \ \sinh
\biggl[ \frac{ 2 \pi T |x-x_1|}{ v_F \sin \alpha} \biggl] }
\nonumber\\
\times J_0 \biggl\{ \frac{2 \lambda}{\sin \alpha}   \sin \biggl[
\frac{ e H d (x-x_1)}{2 c} \biggl] \sin \biggl[ \frac{ e H d
(x+x_1)}{2 c} \biggl] \biggl\}
\nonumber\\
\times \cos \biggl[ \frac{2  \mu_B H (x-x_1)}{v_F \sin \alpha}
\biggl] \ \ \Delta(x_1) \biggl>_{\alpha} ,
\end{eqnarray}
where $g$ is the effective electron coupling constant,
$<...>_{\alpha}$ stands for averaging procedure over angle
$\alpha$. Below, we introduce more convenient variable,
$z=(x-y)/\sin \alpha$. In this case, we can rewrite Eq.(16) in the
following way:
\begin{eqnarray}
\Delta(x) = g \int^{\infty}_{\frac{v_F}{\Omega}} \frac{2 \pi T
dz}{v_F  \sinh \biggl( \frac{ 2 \pi T z}{ v_F} \biggl) } \cos
\biggl( \frac{2  \mu_B H z}{v_F} \biggl)
\nonumber\\
\times \biggl< J_0 \biggl\{ \frac{2 \lambda}{\sin \alpha}   \sin
\biggl( \frac{\omega_c z \sin \alpha}{2 v_F} \biggl) \sin \biggl[
\frac{\omega_c (2x-z \sin \alpha)}{2 v_F} \biggl] \biggl\}
\nonumber\\
 \times \Delta(x - z \sin \alpha) \biggl>_{\alpha} \ .
\end{eqnarray}

Below, we treat the orbital effect against superconductivity as a
small perturbation. To this end, we can expend the Bessel function
in Eq.(17) with respect to small parameter, $\lambda \ll 1$,
(8),(9):
\begin{eqnarray}
&&J_0 \biggl\{ \frac{2 \lambda}{\sin \alpha}   \sin \biggl(
\frac{z \omega_c \sin \alpha}{2 v_F} \biggl) \sin \biggl[
\frac{\omega_c (2x-z \sin \alpha)}{2 v_F} \biggl] \biggl\} \approx
\nonumber\\
&&1 - \frac{\lambda^2}{2 \sin^2\alpha} \sin^2 \biggl(\frac{z
\omega_c \sin \alpha}{2v_F} \biggl)
\nonumber\\
&&+\frac{\lambda^2}{2 \sin^2\alpha} \cos \biggl( \frac{2 \omega_c
x}{v_F} \biggl) \sin^2 \biggl( \frac{z \omega_c \sin \alpha}{2v_F}
\biggl) \cos \biggl( \frac{z \omega_c \sin \alpha}{v_F} \biggl)
\nonumber\\
&&+\frac{\lambda^2}{2 \sin^2\alpha} \sin \biggl( \frac{2 \omega_c
x}{v_F} \biggl) \sin^2 \biggl( \frac{z \omega_c \sin \alpha}{2v_F}
\biggl) \sin \biggl( \frac{z \omega_c \sin \alpha}{v_F} \biggl) .
\end{eqnarray}
It is possible to make sure that Eqs.(17) under the approximation
(18) has the following solution at $T=0$:
\begin{equation}
\Delta(x) = \cos(k_1x) + A \cos(k_1 x) \cos(k_2 x) + B \sin(k_1 x)
\sin(k_2 x) ,
\end{equation}
where $k_1 = 2 \mu_B /v_F$ and $k_2 = 2 \omega_c / v_F$; $A \sim B
\sim \lambda^2$. Note that, in Eqs.(18),(19), we keep only terms
of the order of $\lambda^2$ and disregard all terms of the order
of $\lambda^4$ or less. After substituting Eqs.(18),(19) into
integral Eq.(17) and disregarding all terms of the order of
$\lambda^4$, we obtain the following three equations at $T=0$:
\begin{eqnarray}
\frac{1}{g} = \int^{\infty}_{\frac{v_F}{\Omega}} \frac{dz}{z}
\biggl< \biggl[1 - \frac{\lambda^2}{2 \sin^2 \alpha} \sin^2
\biggl(\frac{k_2 z \sin \alpha}{4} \biggl) \biggl]
\nonumber\\
\times \cos ( k_1 z) \cos(k_1 z \sin \alpha) \biggl>_{\alpha} ,
\end{eqnarray}

\begin{eqnarray}
&&(A+B) \biggl\{ \frac{1}{g} - \int^{\infty}_{\frac{v_F}{\Omega}}
\frac{dz}{z} \cos(k_1z) J_0[(k_1-k_2)z]\biggl\}
\nonumber\\
&&= \frac{\lambda^2}{2} \int^{\infty}_{\frac{v_F}{\Omega}}
\frac{dz}{z} \biggl<\frac{1}{\sin^2 \alpha} \sin^2
\biggl(\frac{k_2 z \sin \alpha}{4}\biggl)
\nonumber\\
&&\times \cos \biggl[\frac{(k_1-k_2)z \sin \alpha}{2} \biggl]
\cos(k_1 z) \biggl] \biggl>_{\alpha} ,
\end{eqnarray}

\begin{eqnarray}
&&(A-B) \biggl\{ \frac{1}{g} - \int^{\infty}_{\frac{v_F}{\Omega}}
\frac{dz}{z} \cos(k_1z) J_0[(k_1+k_2)z]\biggl\}
\nonumber\\
&&= \frac{\lambda^2}{2} \int^{\infty}_{\frac{v_F}{\Omega}}
\frac{dz}{z} \biggl<\frac{1}{\sin^2 \alpha} \sin^2
\biggl(\frac{k_2 z \sin \alpha}{4}\biggl)
\nonumber\\
&&\times \cos \biggl[\frac{(k_1+k_2)z \sin \alpha}{2} \biggl]
\cos(k_1 z) \biggl] \biggl>_{\alpha} .
\end{eqnarray}
Note that Eq.(20) defines correction to the FFLO critical magnetic
field due to the orbital effect at $T=0$, whereas Eqs.(21),(22)
define corrections (19) to the FFLO solution (15).

In this paper, we restrict our analysis by calculation of
correction (20) to the FFLO critical magnetic field, $H_{FFLO}$.
Let us recall that, in the absence of a magnetic field, Eq.(17)
reduces to:
\begin{equation}
\frac{1}{g}= \int^{\infty}_{\frac{v_F}{\Omega}} \frac{2 \pi T_{c0}
dz}{v_F  \sinh \biggl( \frac{ 2 \pi T_{c0} z}{ v_F} \biggl) },
\end{equation}
where $T_{c0}$ is the superconducting transition temperature at
$H=0$. Note that, if orbital effect is negligible (i.e., at
$\lambda =0$), then the FFLO critical magnetic field, $H_{FFLO}$
satisfy the following equation (20):
\begin{equation}
\int^{\infty}_{\frac{v_F}{\Omega}} \frac{2 \pi T_{c0} dz}{v_F
\sinh \biggl( \frac{ 2 \pi T_{c0} z}{ v_F} \biggl) } =
\int^{\infty}_{\frac{v_F}{\Omega}} \frac{dz}{z} \cos ( k_1 z) J_0
(k_1 z ),
\end{equation}
where
\begin{equation}
J_0 (k_1 z ) = <\cos(k_1 z \sin \alpha)>_{\alpha}.
\end{equation}
As shown in Ref.[20], in a pure 2D case, Eq.(24) has the following
solution (see also Ref.[45]):
\begin{equation}
H_{FFLO} = \frac{\Delta_0}{\mu_B} = \frac{\pi k_b T_{c0}}{2
\gamma},
\end{equation}
where $\Delta_0$ is superconducting gap in the
Bardeen-Cooper-Schrieffer theory [1], $k_B$ is the Boltzmann
constant, and $\gamma$ is the Euler constant [45].

Therefore, Eq.(20) can be rewritten as
\begin{eqnarray}
&&\ln (H_{FFLO}/H^*_{FFLO}) =  \lambda^2 \ \int^{\infty}_0
\frac{dz}{z}
\nonumber\\
&&\times \biggl< \frac{\sin^2 (k_2 z \sin \alpha/4)}{2 \sin^2
\alpha}
 \cos ( k_1 z) \cos(k_1 z \sin
\alpha) \biggl>_{\alpha} ,
\end{eqnarray}
where $H^*_{FFLO}$ is critical field of the FFLO phase in the
presence of the orbital effect. In this paper we consider the case
of the small orbital effect (9), thus, Eq.(27) can be rewritten in
the following way:
\begin{eqnarray}
&&(H_{FFLO}-H^*_{FFLO})/H_{FFLO}=  \lambda^2 \ \int^{\infty}_0
\frac{dz}{z}
\nonumber\\
&&\times \biggl< \frac{\sin^2 (k_2 z \sin \alpha/4)}{2 \sin^2
\alpha}
 \cos ( k_1 z) \cos(k_1 z \sin
\alpha) \biggl>_{\alpha}.
\end{eqnarray}
It is possible to make sure that integral in Eqs.(27),(28) is
convergent one. Moreover, the integral is small since it is
proportional to $\lambda^2 \ll 1$ and, thus, the FFLO phase is
stable even in the presence of the orbital effect. As we have
already discussed in the "qualitative" part of this paper,
physically this means that the FFLO superconducting pair is
located mostly within one conducting layer. Under such condition
(9), the intra-layers currents are small and, in fact, we have
coexistence of the FFLO phase [10,11] and the reentrant
superconductivity [4-6].

Let us demonstrate that the above mentioned situation corresponds
to the existence of the FFLO phase in the Q2D superconductor
$\kappa$-(ET)$_2$Cu(NCS)$_2$, where, in our opinion, it is the
most firmly experimentally established [40]. Indeed, if we take
experimental value of the perpendicular upper critical field,
$H^{\perp}_{c2} \simeq 5 \ T$, we obtain the Ginzburg-Landau
parallel coherence length $\xi_{\parallel} \simeq 0.8 \times
10^{-6} cm$ from the standard equation: $H^{\perp}_{c2} = \tau
\phi_0/(2 \pi \xi^2_{\parallel})$, where $\tau=(T_{c0}-T)/T_{c0}$,
$\phi_0$ is the flux quantum. Then, from the equation
$\xi_{\parallel} = \sqrt{7 \zeta(3)} v_F /(4 \sqrt{2} \pi T_{c0})$
[46], we find the value of in-plane Fermi velocity, $v_F \simeq
0.65 \times 10^7 cm/s$. If we take into account that the
inter-plane distance is $d = 1.62 \times 10^{-7} cm$, we obtain
the cyclotron frequency (7): $\omega_c(H)/H \simeq 1.23 \ K/T$.
So, in the integral (28), $k_2/k_1 \simeq 1.85$ and, as it is
possible to show, its numerical evaluation gives the value of
0.12. Therefore, Eq.(28) can be rewritten as
\begin{equation}
H_{FFLO}-H^*_{FFLO} = 0.12 \ \lambda^2.
\end{equation}
Estimation of $t_{\perp} \simeq 2 \ K$ [47], gives us the
following value of parameter $\lambda = 0.16$ in the vicinity of
the magnetic field $H_{FFLO} \simeq 27.5 \ T$. So, we can conclude
that indeed, in the Q2D superconductor
$\kappa$-(ET)$_2$Cu(NCS)$_2$, the FFLO phase coexists with the
reentrant superconductivity. Note that qualitatively the above
mentioned statement does not depend on actual symmetry of
superconducting gap, which may be $d$-wave in the
$\kappa$-(ET)$_2$Cu(NCS)$_2$.

Let us consider another relative Q2D organic conductor -
$\kappa$-(ET)$_2$Cu[N(CN)$_2$]Cl (see, for example, Ref. [31]). In
accordance with [31], in this case under pressure $P = 1.9 \
kbar$, $T_{c0} \simeq 7 \ K$ and $H^{\perp}_{c2} \simeq 2 \ T$.
Using the same equations as before, we obtain $\xi_{\parallel}
\simeq 1.25 \times 10^{-6} cm$ and $v_F \simeq 0.73 \times 10^{7}
cm/s$. Moreover, from Ref.[31], it follows that
$H^{\parallel}_{c2} \simeq 20 \ T$ and, using equation
$\xi_{\perp} = \sqrt{7 \zeta(3)} 2 t_{\perp} d /(4 \sqrt{2} \pi
T_{c0})$ [46], we obtain $t_{\perp} \simeq 17 \ K$. Taking into
account that $d \simeq 1.5 \times 10^{-7} cm$, we find that, in
this case, the parameter $\lambda \simeq 2.6$ is large and Eq.(18)
is not valid. In other words, the orbital effect against
superconductivity is important and, thus, it is necessary to solve
Eq.(17) directly for $\lambda \geq 1$. However, this is beyond the
scope of the current paper.

To summarize, we have shown that, at small values of the parameter
$\lambda \ll 1$ in Eq.(8), the superconducting FFLO phase in a
parallel magnetic field occupies almost one conducting layer at
$T=0$. In this case, the FFLO phase [10,11] exists under the
reentrant superconductivity regime [4-6] and the correction from
the orbital effect to the FFLO critical magnetic field (28) is
small. Such situation has been shown to exist in the Q2D
superconductor $\kappa$-(ET)$_2$Cu(NCS)$_2$. If parameter
$\lambda$ is of the order of unity, as it is in the case of
another Q2D organic conductors $\kappa$-(ET)$_2$Cu[N(CN)$_2$]Cl,
then the orbital effect becomes large and Eq.(17) needs to be
solved without expanding the Bessel function. The latter problem
is very difficult from numerical point of view and hopefully will
be considered in the future. We stress that our results are
different from that in Refs.[27,28], since at zero temperature it
is not possible to expand the superconducting gap equation with
respect to parameter $t_{\perp}/T$. In the end of the paper, we
discuss in a brief one delicate property of our model - that the
direction of the FFLO phase is supposed to be unchanged in a
magnetic field. This definitely works for the case of small
magnetic fields, considered in the paper, where anisotropy of the
2D FS fixes the FFLO direction. As to relatively high magnetic
fields, the effect of changing of the FFLO direction has to be
somehow added to Eq.(17).

We are thankful to C.C. Agosta, N.N. Bagmet (Lebed), and M.V.
Kartsovnik for useful discussions.

$^*$Also at: L.D. Landau Institute for Theoretical Physics, RAS, 2
Kosygina Street, Moscow 117334, Russia.

\end{document}